\documentclass[aps, prl,twocolumn,superscriptaddress,10pt]{revtex4-1}

\usepackage{comment}
\usepackage{graphicx}
\graphicspath{ {./pics/} }
\usepackage{amsmath}

\usepackage[locale=US,mode=math]{siunitx}
    \DeclareSIUnit\gravity{g}
    \DeclareSIUnit\grms{Grms}

\usepackage{setspace}
\usepackage{braket}
\usepackage{mathtools}
\usepackage{amssymb}

\usepackage{natbib,mathrsfs,accents}
\usepackage{url} %for url ref in bib
\usepackage[dvipsnames]{xcolor}
\definecolor{myblue}{named}{MidnightBlue}
\definecolor{mygreen}{RGB}{0,120,0}
\usepackage[caption=false]{subfig}
\usepackage{float}

\begin{document}

\title{%A 698 nm laser system for excitation of hexagonal Boron Nitride (hBN) based single photon source on a CubeSat mission
A 698 nm laser system for excitation of fluorescent quantum light sources on a CubeSat mission }


\author{Sven Schwertfeger}
\affiliation{Ferdinand-Braun-Institut (FBH), Gustav-Kirchoff-Str.4, 12489 Berlin}

\author{Elisa Da Ros}
\email{elisa.da.ros@physik.hu-berlin.de}
\affiliation{Institut f\"{u}r Physik and CSMB, Humboldt-Universit\"{a}t zu Berlin, Newtonstr. 15, Berlin 12489, Germany}

\author{Marcel Bursy}
\affiliation{Ferdinand-Braun-Institut (FBH), Gustav-Kirchoff-Str.4, 12489 Berlin}

\author{Andreas Wicht}
\affiliation{Ferdinand-Braun-Institut (FBH), Gustav-Kirchoff-Str.4, 12489 Berlin}

\author{Daniel Pardo}
\affiliation{Centre for Quantum Technologies, Department of Physics, National University of Singapore, 117543 Singapore, Singapore}

\author{Ankush Sharma }
\affiliation{Centre for Quantum Technologies, Department of Physics, National University of Singapore, 117543 Singapore, Singapore}

\author{Subash Sachidananda}
\affiliation{Centre for Quantum Technologies, Department of Physics, National University of Singapore, 117543 Singapore, Singapore}

\author{Alexander Ling}
\affiliation{Centre for Quantum Technologies, Department of Physics, National University of Singapore, 117543 Singapore, Singapore}

\author{Markus Krutzik} 
\affiliation{Ferdinand-Braun-Institut (FBH), Gustav-Kirchoff-Str.4, 12489 Berlin}
\affiliation{Institut f\"{u}r Physik and CSMB, Humboldt-Universit\"{a}t zu Berlin, Newtonstr. 15, Berlin 12489, Germany}

\begin{abstract}{
This manuscript reports on the development and qualification of an ECDL-based, fiber-coupled laser system at a wavelength of $\lambda = \SI{698}{\nano\meter}$ for space applications. We designed and developed the optical and mechanical configuration, along with the laser driving and thermal management electronics, to meet space compatibility requirements. Validation tests were conducted on off-the-shelf components to assess their suitability for satellite deployment. The final system integrates all components into a compact design optimized for CubeSat platforms. 
}
\end{abstract}

%\begin{keyword}
%ECDL, Laser, space test, new space
%\end{keyword}
\maketitle
%\end{frontmatter}
\section{Introduction}

Satellite-based quantum technologies are essential for advancing both applied and fundamental science. In recent years, numerous proposals have highlighted the potential of quantum technologies in space and have shown roadmaps towards their deployment \cite{Tino2013, Schuldt2015, Aguilera2014, Gurlebeck2018, El-Neaj2020, Kaltenbaek2021, Sidhu2021, Schkolnik2023, Orsucci2025, Gaaloul2025}.

For implementation on satellites, components must typically meet stringent requirements for radiation hardness, environmental resilience, and operational reliability throughout the mission's lifetime \cite{Nasa-Space1}. Furthermore, the conditions on a CubeSat platform impose strict constraints on size, weight, and power (SWaP)\cite{Nasa-Space2,Strangfeld2021}.

In this work, we present an ECDL-based fiber-coupled laser unit developed for space application using the new space philosophy \cite{NewSpace}. This approach forgoes the traditional full space certification, in favor of testing conventional off-the-shelf components under conditions that simulate the launch and space environment and relying on space heritage, i.e. 
using technologies and concepts previously tested through missions on sounding rockets \cite{Lezius16, Schkolnik2016, Maius2018, Doringshoff2019, Adams2020, Probster2021} and small satellites, as \cite{SpooQy, Oi2017}. Although this entails higher risks, it offers advantages in terms of cost reductions and faster development cycles.

\begin{figure}[h]
  \includegraphics[width=8.5cm]{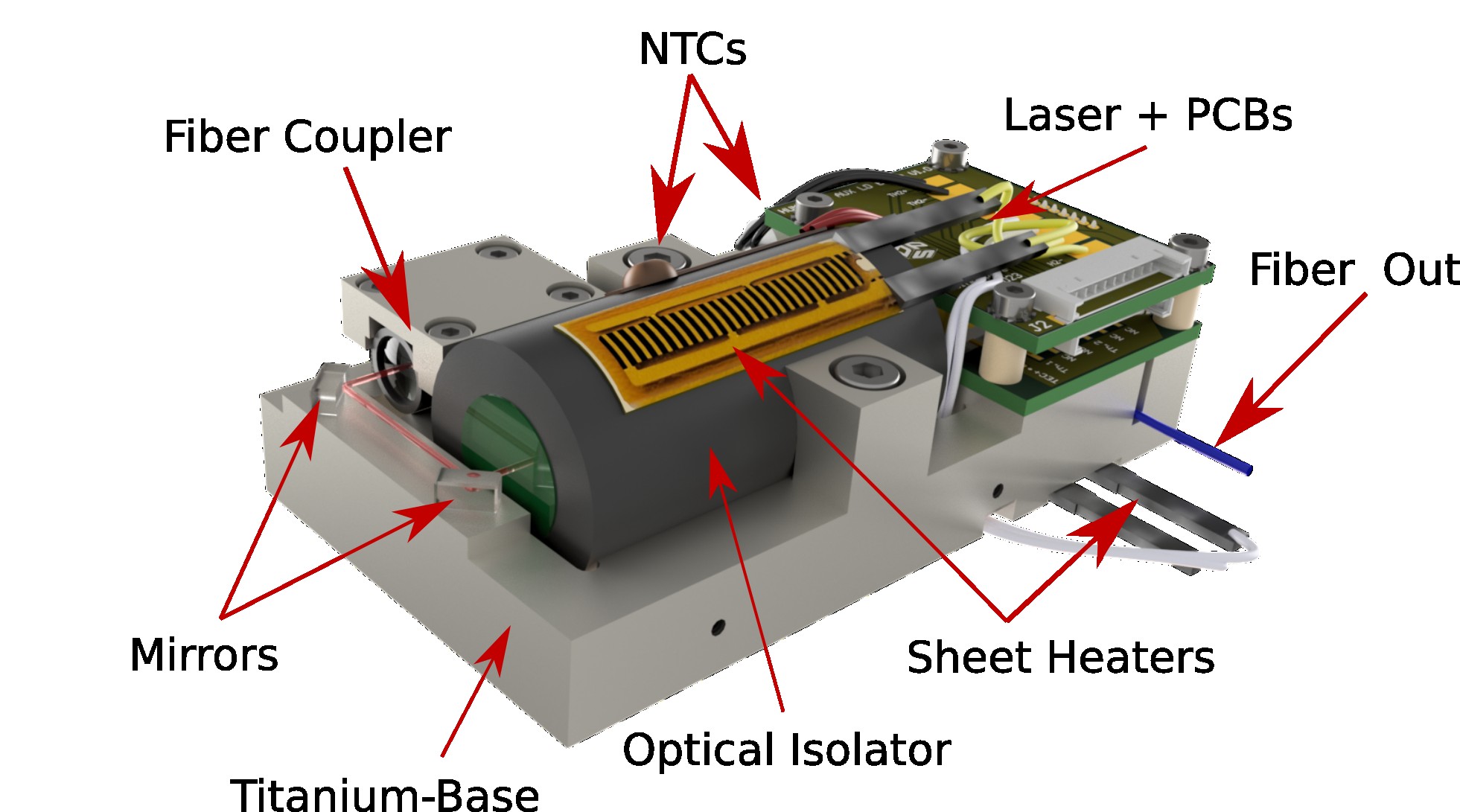}
    \caption{CAD rendering of the laser unit, excluding any housing and laser driving electronics}
    \label{fig:DesignLaserModul}
\end{figure}

We report on the development of the mechanical, optical, and electronic design of the laser unit, as well as the benchmarking of the chosen components with respect to their ability to withstand mechanical vibrations, pyrotechnic events, thermal vacuum cycles and irradiation.
The laser unit operates at the wavelength of \SI{698}{\nano\meter}, which poses an unique challenge: to our knowledge, this wavelength has not been previously investigated or implemented in the context of space applications. Despite this, it has been specifically chosen to excite single-photon sources (SPS) based on color centers in the two-dimensional material hexagonal boron nitride (hBN)~\cite{Cholsuk2024}, as represented in Fig.~\ref{fig:DesignLaserModul}. 
This laser unit is developed within the QUICK$^{3}$ project (QUantum phonIsChe Komponenten für sichere Kommunikation mit Kleinsatelliten)\cite{Quick3HP}. 
The aim of the QUICK$^{3}$ mission is to design an SPS based on a fluorescent defect in hBN and evaluate its functionality in space, deploying it on a 3U nano-satellite bus from NanoAvionics\cite{tobi-sps, CubeSat, Ahmadi_2024}. This constitutes the first use of such a design for an hBN-based single-photon source in space. 
The \SI{698}{\nano\meter} wavelength is not only suitable for exciting different color centers in hBN, but, additionally, it corresponds to the strontium optical clock transition~\cite{ Aeppli2024}, highlighting the potential of this technology to enable future space-compatible optical clocks. Developing such devices could significantly enhance timekeeping precision~\cite{Dimarcq2024, Yang2025}, improve global positioning and navigation~\cite{Origlia2018}, as well as enable tests of fundamental physics~\cite{Derevianko2014, Kolkowitz2016, Schkolnik2023}.

\section{Laser System design}
Developing a laser unit for space applications requires careful consideration of its optical, mechanical, thermal and electronic design. 

\subsection{Optical design}
\begin{figure}[ht]
  \includegraphics[width=8cm]{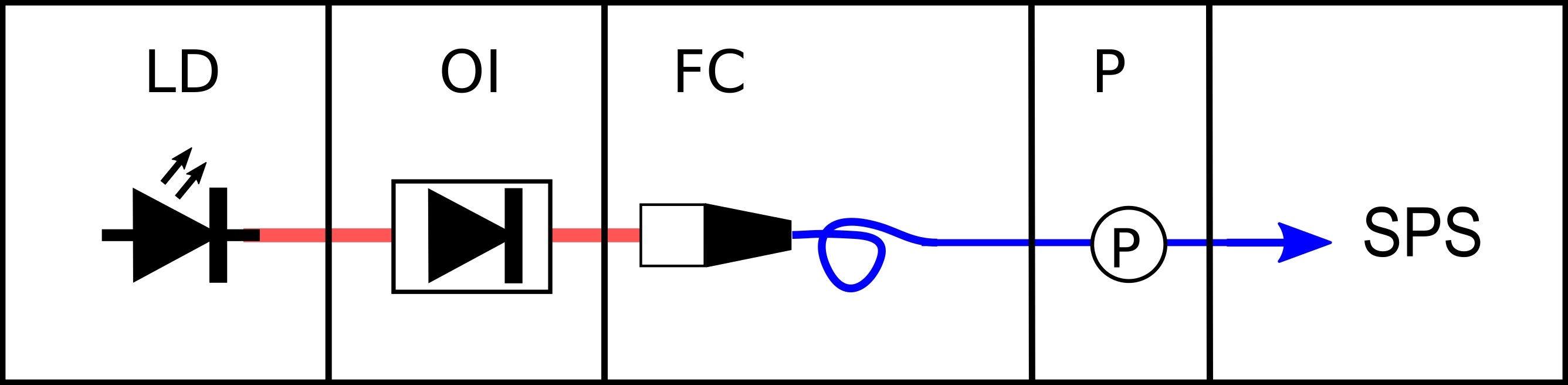}
    \caption{Sketch of the laser system layout.  LD = laser diode, OI = optical isolator, FC = fiber coupler, P = inline powermeter, SPS = SinglePhotonSource\cite{Ahmadi_2024} }
    \label{fig:SimpleApproach}
\end{figure}

The QUICK$^{3}$ optical design adopts a straightforward and efficient approach, coupling a laser diode to an optical fiber via an isolator and mirrors. Figure \ref{fig:SimpleApproach} illustrates the schematic diagram of the laser system layout.

The \SI{698}{\nano\meter} pump laser is a commercially available extended cavity diode laser (ECDL) incorporating a ridge waveguide design housed in a butterfly package, that includes an anti-reflection coated window, a thermoelectric cooler (TEC), and a \SI{10}{\kilo\ohm} NTC thermistor and was hermetically sealed. Optical feedback is provided by an external volume Bragg grating (VHBG) at the front facet. Two cylindrical lenses collimate the output beam to a diameter of around \SI{800}{\micro\meter}. 
The laser delivers up to \SI{30}{\milli\watt} of free-space optical power at \SI{65}{\milli\ampere}.
To deliver laser light to the experiment, we employ a polarization-maintaining (PM) fiber with an integrated coupling lens and an internal Si-photodiode for power monitoring. The tap ratio was set to \SI{4}{\percent}.  We choose adhesive integration of two mirrors for beam steering into the optical fiber. The pre-assembled fiber coupler simplifies the alignment and integration, but it still requires precise control of the collimated beam's angle relative to the ferrule. 

Here, simulations indicate that a misalignment of \SI{0.35}{\milli\radian} would reduce the coupling efficiency by half, whereas positional offsets of up to \SI{100}{\micro\meter} maintain efficiencies above \SI{90}{\percent}. To compensate for the \SI{45}{\degree} polarization rotation by the optical isolator, a $\lambda$/2 waveplate is inserted before the fiber-coupling mirrors, ensuring proper polarization alignment with their reflectivity properties.

\subsection{Mechanical and thermal design}

The laser unit (LU) is mounted on a titanium base, chosen for its trade-off between stiffness, weight, and thermal stability, making it suitable for space applications \cite{BOYER1996103}. All components, including sheet heaters, NTCs, and mirrors, are bonded with appropriate space-qualified adhesives. We used two specialized epoxies, one with a high thermal conductivity, to bond the sheet heaters, and another to secure all other components, such as wires and screws. The mirrors were actively aligned and glued using a UV-curing adhesive. 
First, the laser is mounted on the base, then the ferrule is aligned with respect to the polarization axis with the help of backside-coupled laser light and a polarisation beam splitter (PBS), and fixed with UV-curing adhesive. The two circuit boards that interface with the driver electronics are then installed. The isolator, with an adjustable $\lambda$/2 waveplate, is positioned using the collimated output of the butterfly laser, optimizing the isolator angle for maximum transmission. 
Finally, both mirrors are actively aligned with micro-manipulators \cite{10.1117/12.2253655} and secured with UV-cured adhesive.

Space-qualified connectors ensure the electrical connections between the laser unit and the electrical driver board. In the same fashion, the wiring for the NTCs and sheet heaters adheres to the European Cooperation for Space Standardization guidelines \cite{ecss}. 

\subsection{Electronic design} \label{electronic}

The laser driver, implemented on a single printed circuit board, can be controlled by a satellite main flight controller through a serial interface. The system is designed around a Programmable-System-On-Chip (PSoC3) microcontroller \cite{Cheng15}, along with other off-the-shelf components that have already demonstrated their reliability in past satellite missions like SpooQy-1 \cite{SpooQy}. The driver operates the laser diode in continuous-wave, constant-current mode, providing temperature control, readout of the inline photodetector, and onboard storage via a NAND flash memory. The laser current control and diode protection mechanisms have been explained in \cite{Ahmadi_2024}.

The temperature readout of the laser unit is facilitated by connecting the NTCs to one arm of an unbalanced Wheatstone bridge, where the differential voltage is read out using a differential amplifier connected to a 24-bit Analog-to-Digital Converter. This allows temperature accuracy up to \SI{\pm 2}{\milli\kelvin}. This measurement is used by software-based PID controller to regulate the TEC and sheet heaters.

The TEC is powered by a switch-mode driver, controlled by a 12-bit digital-to-analog converter attached to the PSoC3, allowing current control with a \SI{1}{\milli\ampere} resolution. The driver also allows for over-current and over-voltage protection, along with inline current monitoring. A pulse width modulated (PWM) signal is generated by the PSoC and sent to the gates of two MOSFETs, which in turn control the power delivery to two \SI{6}{\watt} sheet heaters.

\section{Qualification on component level}

This section presents the qualification tests performed to ensure the suitability of all components for our use case. 
Our mission adopts a New Space approach, which accepts higher risks in exchange for reduced qualification timelines, with agile testing procedures and use of commercial off-the-shelf components to enhance cost efficiency. In line with this philosophy, we considered only three units per component: two designated for testing, and one untested component reserved for integration within the flight laser unit.   
Accordingly, two units each of the following commercially available components were tested for resistance to radiation, temperature fluctuations and mechanical stress: the ECDL lasers, the isolators and the PM fibers.

\textit{Random vibration} tests were carried out at stress levels up to $\SI{12}{G_{rms}}$, within a frequency band from \SI{20}{\hertz} to \SI{2}{\kilo\hertz}. Given the system's intended deployment on a Falcon 9, the test parameters were set according to the Falcon standards~\cite{SpaceX}. To this end, all the components were mounted on a test platform and subjected to controlled random vibrations with a well-defined amplitude spectrum for durations ranging from a few seconds up to 8 minutes. This stress test was carried out for all three axes consecutively. Following the random vibration tests, mechanical shock tests were conducted again along all three axes consecutively, each with a peak acceleration of approximately \SI{1000}{\gravity}, and frequencies up to \SI{10}{\kilo\hertz}.

\begin{figure}[h]
  \includegraphics[width=7.5cm]{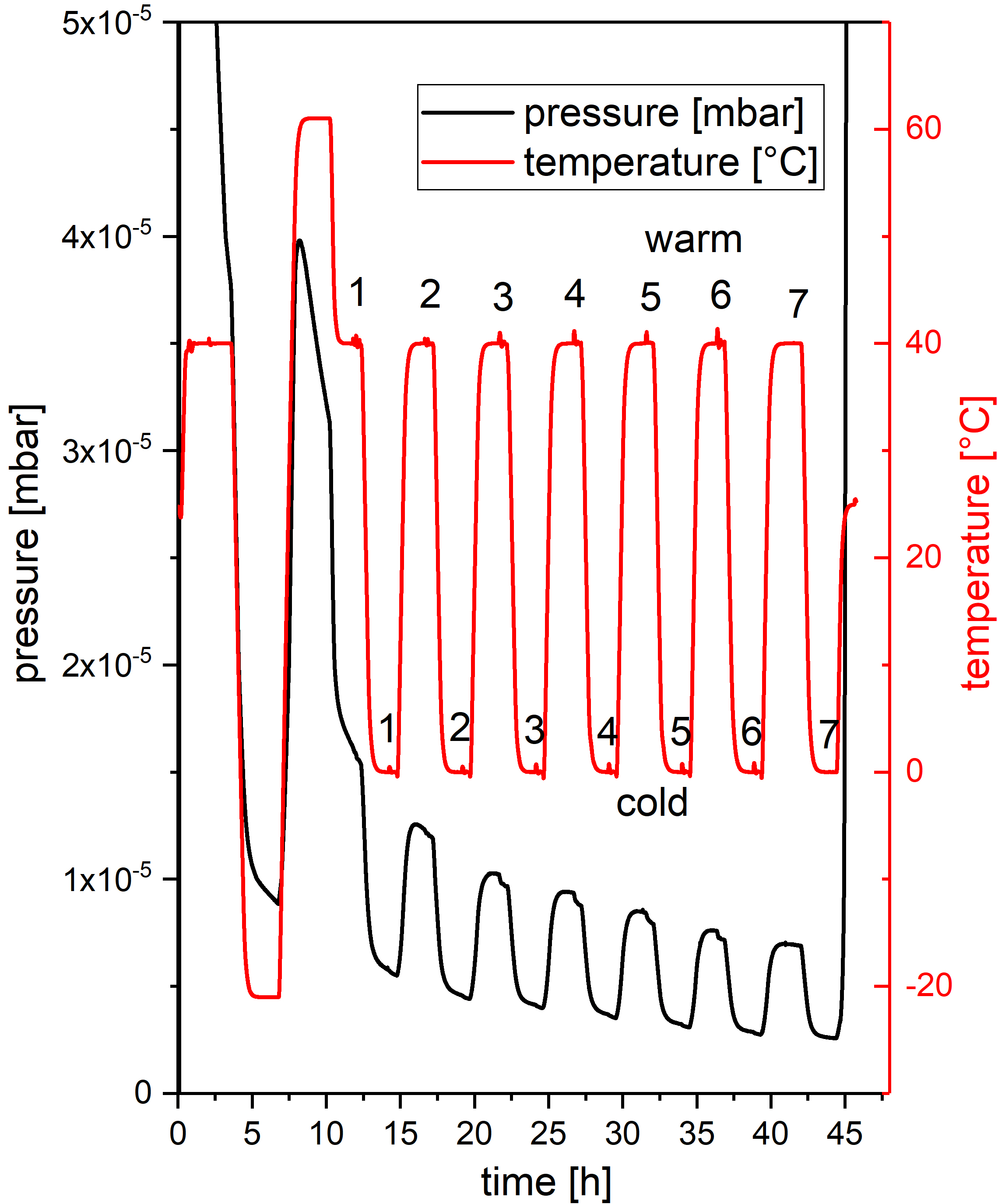}
    \caption{Thermal vacuum test time scheme, measured values of the temperature (red) and pressure (black) inside the test chamber}
    \label{fig:TVTestScheme}
\end{figure}

To simulate \textit{thermal stress} conditions, all components were enclosed in a large vacuum chamber featuring a temperature-controlled plate at the bottom. The isolators and fibers were secured directly on this plate with Kapton tape, while the butterfly lasers were mounted onto an aluminum plate. A temperature sensor, used for regulating the temperature, was installed near one of the butterfly lasers. Fig. \ref{fig:TVTestScheme} shows the trends of pressure and temperature inside the test chamber. The test lasted approximately 45 hours.

During the first three hours, the chamber was evacuated to remove potential residues from the test objects. After the evacuation stage, the test procedure consisted of two phases, a passive phase followed by an active one. During the passive phase, the chamber temperature was lowered to \SI{-20}{\celsius} and subsequently raised to \SI{60}{\celsius}, while the lasers remained off. The active phase comprised seven cycles, spanning a temperature range between \SI{0}{\celsius} and \SI{40}{\celsius}, with each cycle lasting 1.5 hours. In the last 30 minutes of each cycle, the lasers were powered on, and their internal temperature was stabilized at \SI{25}{\celsius} using the integrated TECs. The emitted beams were directed through a window onto a power detector, ensuring continuous output of the laser diode during the test.

The \textit{irradiation} of all components was carried out using a Cobalt-60 source \cite{PTB}. The radioactive decay of $^{\mathrm{60}}$Co releases one electron with an energy of \SI{317.9}{\kilo\electronvolt} and two gamma photons with energies of \SI{1.173}{\mega\electronvolt} and \SI{1.332}{\mega\electronvolt}. The components were exposed for \SI{20}{\hour} at a dose rate of \SI{10}{\gray/ \hour}, resulting in a total absorbed dose of \SI{200}{\gray} for each element. This dose is about 50 times higher than the radiation exposure expected over one year in orbit for the given mission.

\subsection{Test Results}

\begin{figure*}[ht] 
    \centering
    \begin{minipage}[b]{0.48\textwidth}
        \centering
        \includegraphics[width=\textwidth]{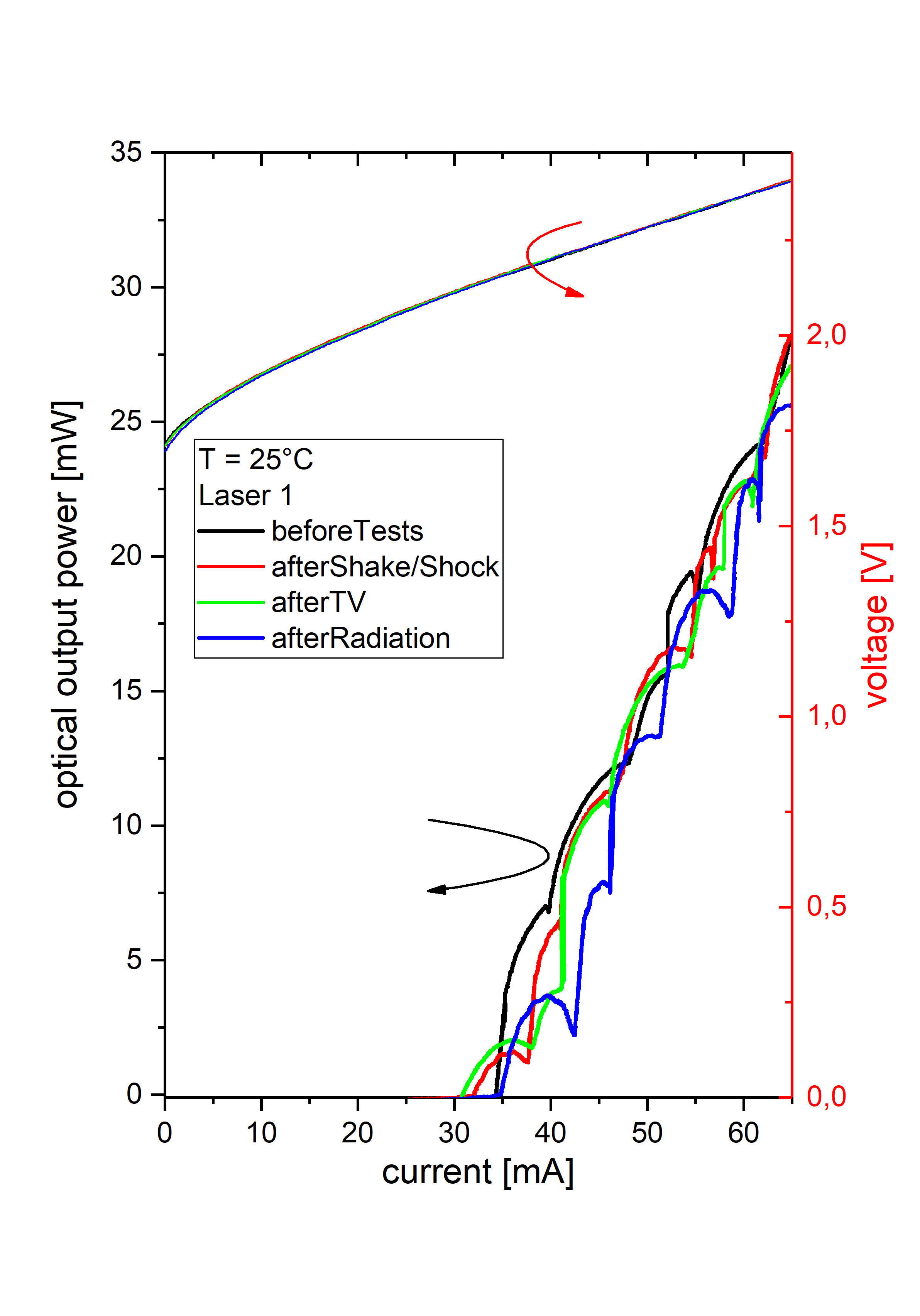}
    \end{minipage}
    \hfill
    \begin{minipage}[b]{0.48\textwidth}
        \centering
        \includegraphics[width=\textwidth]{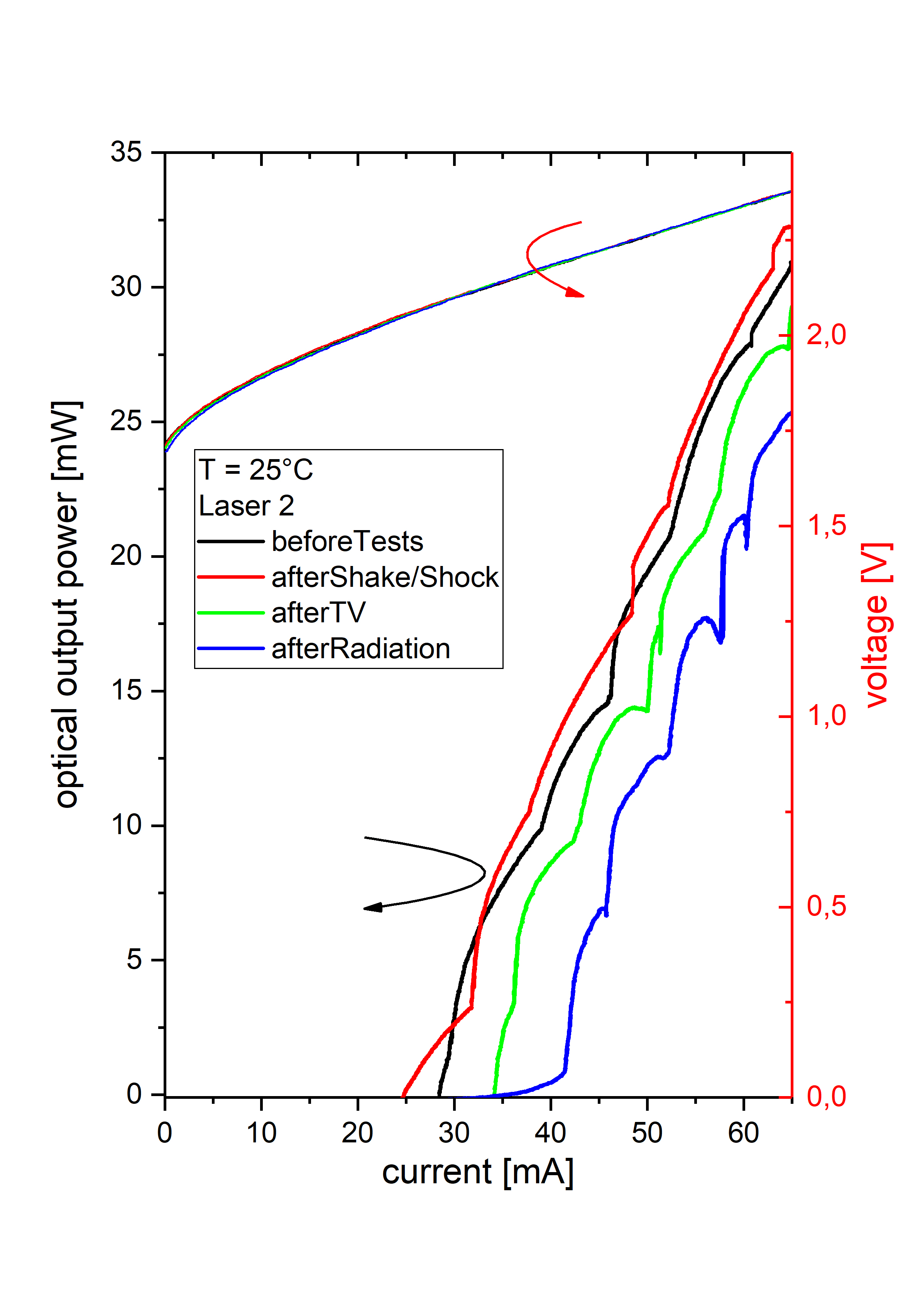}
       
    \end{minipage}
    \caption{Power-current curves for the two tested lasers, before and after each test. The associated voltage is shown by the curves identified by the red arrow }
    \label{fig:PI-Laser}
\end{figure*}

\begin{figure}[h]
  \includegraphics[width=7.5cm]{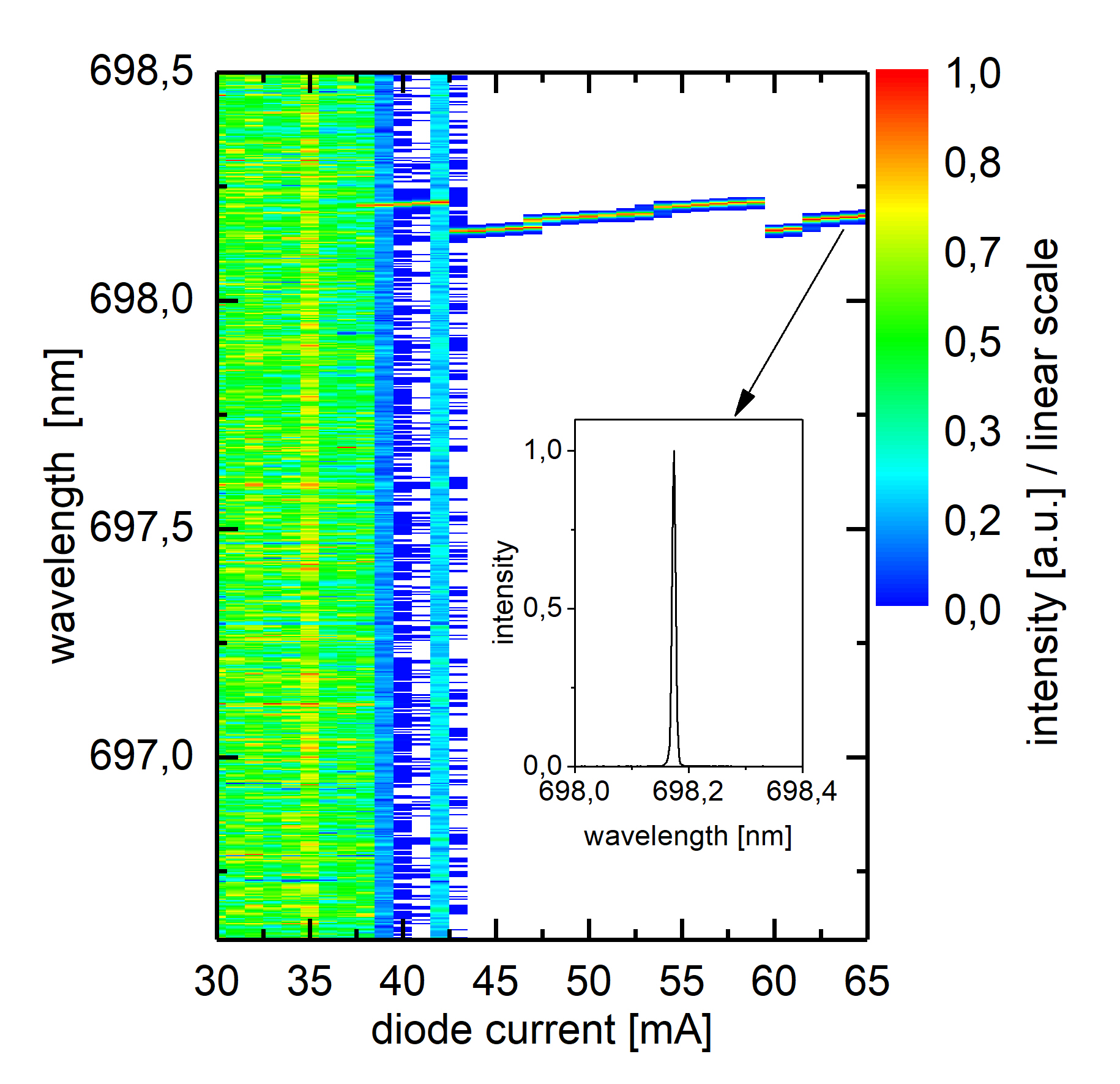}
    \caption{Measured spectra versus current for laser 2 after testing. The normalized spectra are presented as a color map, inset shows single spectrum at I = \SI{63}{\milli\ampere}}
    \label{fig:spectra-Laser2}
\end{figure}

We first examine the response of two butterfly-housed laser diodes to the qualification tests. Power–current characteristics and optical spectra versus current were recorded before and after each test. Figure \ref{fig:PI-Laser} displays the power–current curves for both devices. During the measurements, the laser temperature was stabilized at \SI{25}{\celsius} via the internal TEC.

The power–current curves exhibit minor deviations post-test for both lasers, with laser 2 showing a more pronounced decrease in output power. The optical spectra were measured for both lasers after the test. Given the more significant deviation exhibited in the P-I curves of laser 2, only its spectra are shown in Fig. \ref{fig:spectra-Laser2}, demonstrating single-frequency operation near $\lambda = \SI{698}{\nano\meter}$ above threshold, albeit with occasional mode hops. The inset illustrates a representative spectrum at the diode current of $\sim$\SI{63}{\milli\ampere}. The linewidth, determined using an LTB echelle spectrometer, is below \SI{10}{\pico\meter}, limited by the resolution of the instrument. Similar results were measured on Laser 1.
The spectral map and power–current characteristics after each test remain consistent with the behavior of an extended-cavity diode laser as observed before testing. The reduction in output power for laser 2 suggests a slight misalignment of the optical components within the butterfly housing, potentially aggravated by radiation-induced degradation of the adhesive used to secure the optics. Ultimately, both lasers achieve a maximum output power of approximately \SI{25}{\milli\watt} after testing, with output spectra well defined by the internal grating. The observed degradation does not compromise the mission compatibility, as the reduced output still significantly exceeds the power required, which is below \SI{1}{mW} to pump the SPS. Additionally, the TEC and NTC responsible for temperature control inside the butterfly housing perform as effective as before the tests.

The insertion loss and isolation of the optical isolators were measured before and after each test, showing no detectable changes within the measurement accuracy. Isolation remains above \SI{30}{dB}, while insertion loss consistently stays around \SI{0.5}{dB}.

Similarly, fiber measurements showed no variations in the coupling efficiency or inline photodiode sensitivity. The coupling efficiency exceeds \SI{55}{\percent}, and the sensitivity of the inline photodiode remains approximately \SI{13.8}{\milli\ampere\per\watt}.

\section{Assembled Laser Unit}

\begin{figure}[h]
  \includegraphics[width=7.5cm]{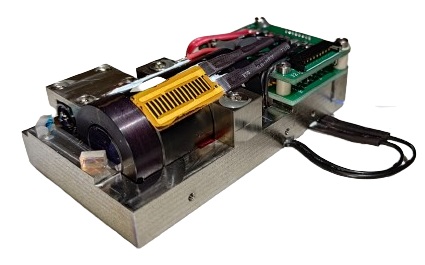}
    \caption{Fully functional engineering qualification model of the laser unit without the lid }
    \label{fig:picQLU}
\end{figure}

A physical mock-up of the LU was assembled to assess the feasibility of the design and the practical steps of the integration process. The laser integrated into the mock-up is the unit designated as laser 2 during testing. After successfully completing the assembly and optical characterization, a second laser unit, meant as the engineering qualification model (EQM) was constructed, as shown in Fig. \ref{fig:picQLU}. An aluminum lid, produced via additive manufacturing, was also created to protect the laser unit. The EQM LU was built using a previously tested isolator and fiber, paired with an untested laser from the same production batch of the tested ones. We could then proceed with a full characterization of the EQM LU.

With an insertion loss of approximately \SI{1}{\percent} in the isolator and a measured coupling efficiency into the fiber of around \SI{50}{\percent}, the maximum fiber-coupled power output of the unit reaches about \SI{13}{\milli\watt} at \SI{698.4}{\nano\meter}. 

\begin{figure}[h]
  \includegraphics[width=7.5cm]{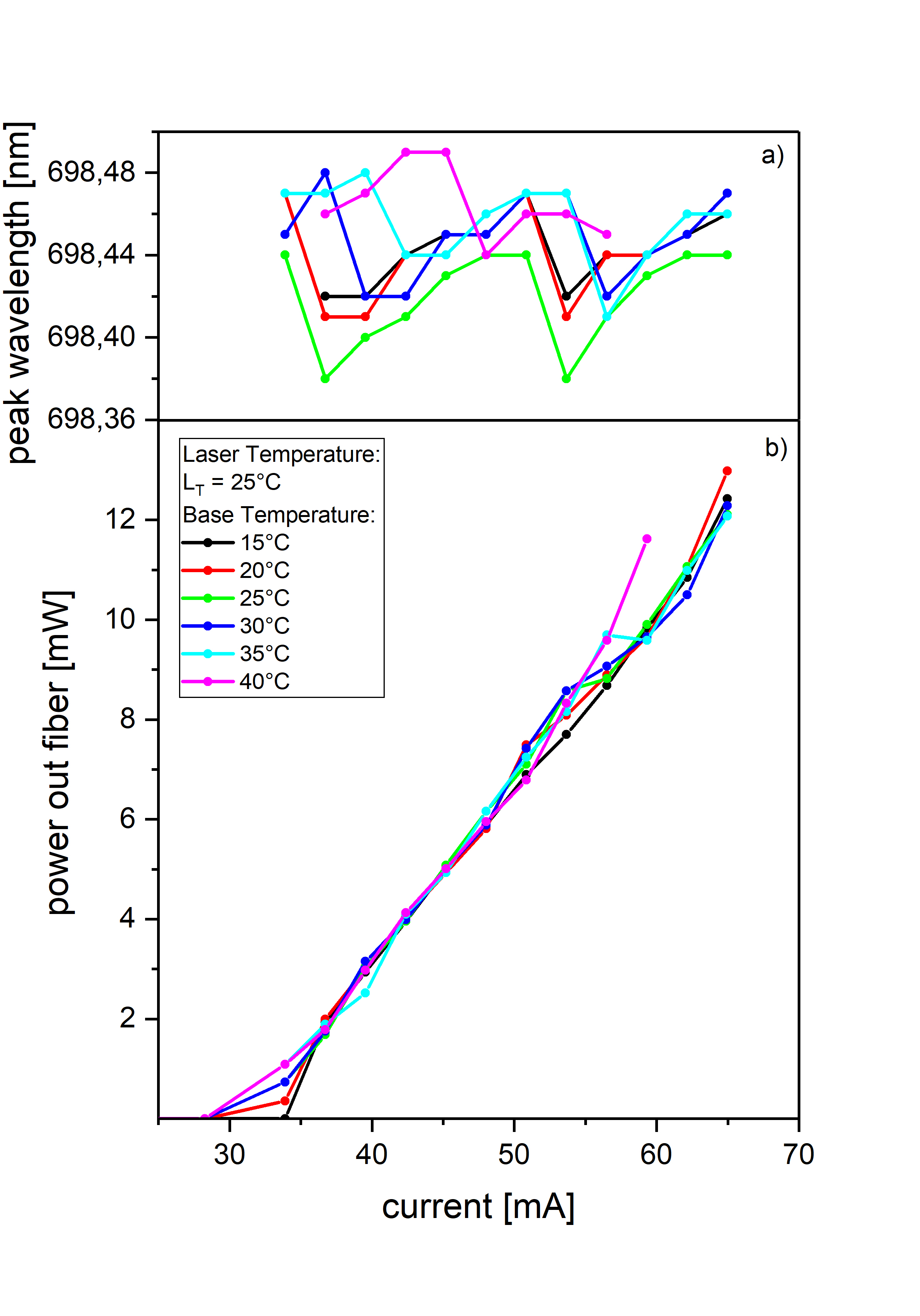}
    \caption{a) Peak values of the spectra for different temperatures and output power above threshold b) Fiber-coupled power output above threshold of the EQM  for different temperatures of the titanium base }
    \label{fig:PUI-spare}
\end{figure}

The power-current characteristics of the laser unit were measured for different temperatures, as shown in Fig. \ref{fig:PUI-spare} b). The temperature of the titanium base was varied between \SI{15}{\celsius} and \SI{40}{\celsius} using an external heater, while the laser diode itself was stabilized at \SI{25}{\celsius} with the internal TEC.

The power output from the fiber was measured simultaneously through an integrating sphere and the fiber inline power detector. The results show that varying environmental temperatures have no significant impact on the laser's power output.

The spectral emission behavior was also examined as a function of the laser diode currents at different temperatures of the base. %above threshold. 
The laser temperature was fixed at \SI{25}{\celsius} via the internal TEC, while the base temperature was varied from \SI{15}{\celsius} to \SI{40}{\celsius}. Under all conditions, single-mode emission - similar to the inset of Fig. \ref{fig:spectra-Laser2} - was observed. On top of figure \ref{fig:PUI-spare} we presents the peak emission wavelength for different diode currents and base temperatures. Due to typical mode hops in ECDL lasers \cite{WOS:000463026500002}, the peak wavelength fluctuated within a range of \SI{0.12}{\nano\meter} throughout all measurements.

\section{Laser unit with micro-isolator}

The commercial optical isolator characterized in the previous sections produces a strong magnetic field that is incompatible with the constraints for a 3U CubeSat, such as the one considered for this work. In particular, the field strength is sufficient to interfere with the satellite's magnetic field sensors. Additionally, the magnetic field would cause the satellite to align with Earth's magnetic field, regardless of any compensation from the reaction wheels and magnetorquers, potentially rendering the mission inoperable.

To address this issue, an alternative component is implemented, relying on a micro-isolator based on a CdMnTe Faraday-crystal \cite{WOS:000914852700004, bursy_isolator}, in combination with a magnetic shield. This micro-isolator emits a magnetic field that is approximately 4-fold weaker than the commercial device. Furthermore, its compact dimensions (\SI{6.5}{}$\times$\SI{6.5}{}$\times$\SI{10}{})~\SI{}{\milli\cubic\metre}, compared to the commercial counterpart of size $\varnothing$\SI{22}{}$\times$\SI{40}{\milli\cubic\metre}, allow the implementation of a cylindrical magnetic shield without increasing the volume of the laser unit.
This micro-isolator has also successfully passed the same qualification tests outlined in the previous section, making it well-suited for our application. In addition, its design for space applications substantially mitigates the associated mission risks.

\begin{figure}
    \centering
    \includegraphics[width=0.9\linewidth]{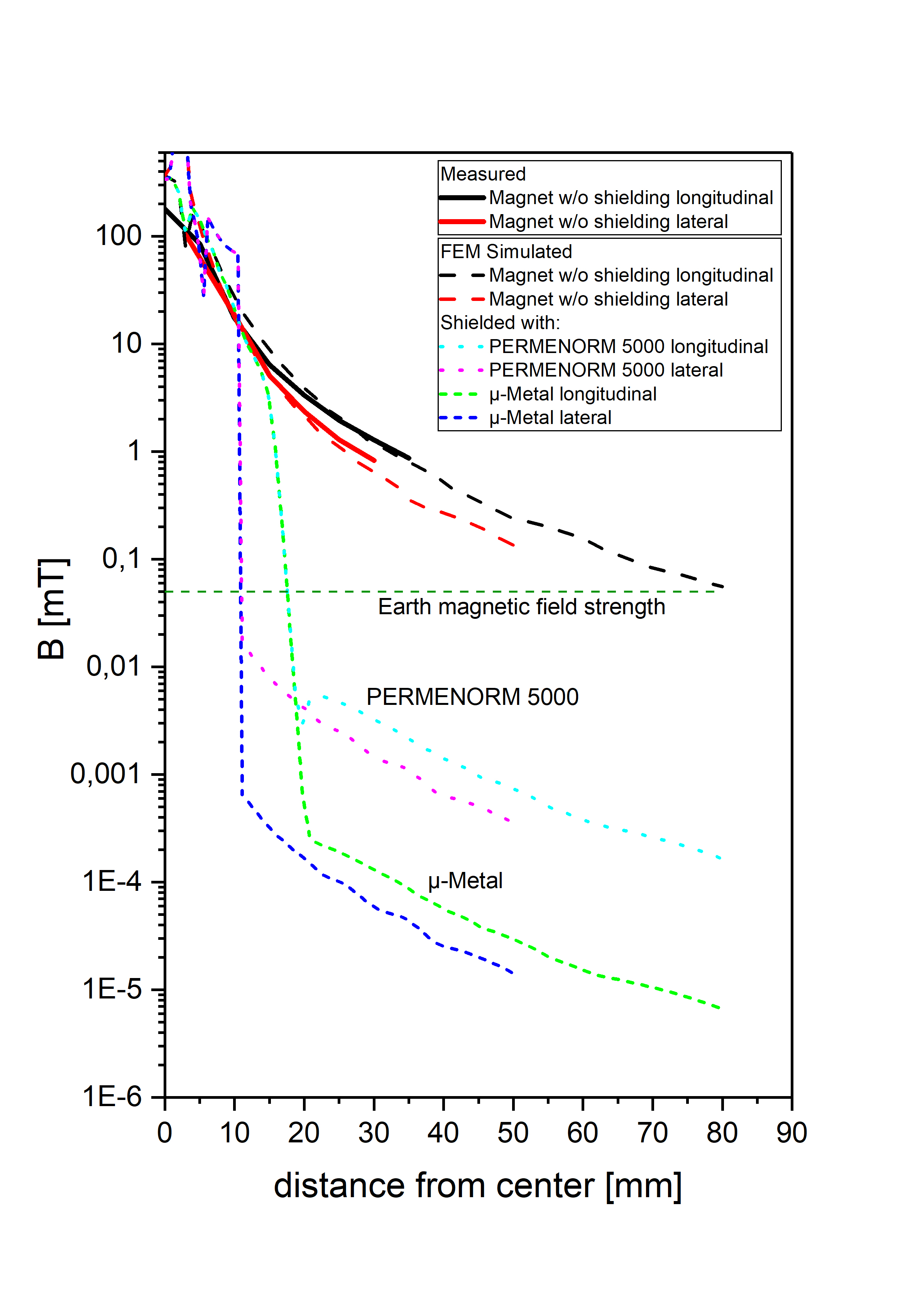}
    \caption{Simulation of the magnetic field strength of the micro-isolator with various magnetic shielding configurations, compared to both measured and simulated field strengths in the absence of shielding.}
    \label{fig:magnetsimu}
\end{figure}

To ensure compatibility with the 3U CubeSat planned for this mission, the stray magnetic field generated by the micro-isolator cannot exceed the strength of Earth's geomagnetic field. However, as shown in Fig. \ref{fig:magnetsimu}, the magnetic field produced by the micro-isolator, although it is lower than the one of the commercial isolator, it already violates this constraint.

The magnetic field has been characterized in both longitudinal and lateral directions using a Hall sensor in an unshielded laboratory environment. Due to the sensor's sensitivity limit of \SI{1}{\milli\tesla}, the measurements were collected as a function of the distance from the isolator's center until the magnetic field dropped below the detection threshold. 

A magnetic shield was designed in response to the satellite’s magnetic field requirement, with its performance evaluated via simulations across a range of wall thicknesses and materials. 
Each simulation was performed using finite element method (FEM) via the Ansys software. This allowed the creation of a digital twin of the isolator's magnet that accurately reproduced the measured magnetic field and enabled reliable extrapolation to larger distances. This virtual model was then used to design a cylindrical magnetic shield matching the dimensions of the commercial isolator, allowing for direct substitution in the optical layout without mechanical modifications.
Figure~\ref{fig:magnetsimu} highlights the comparisons between the unshielded isolator and shields with a wall thickness of \SI{5}{milli}{meter} made of $\mathrm{\mu}$-metal and PERMENORM~5000.
Both $\mathrm{\mu}$-metal and PERMENORM~5000 enable an external field attenuation below the ambient geomagnetic field. Although $\mathrm{\mu}$-metal provided slightly better attenuation and a higher resilience to vibration and shocks, PERMENORM~5000 was selected due to its higher saturation threshold and higher operational stability, which is necessary for our mission.

Upon integrating the isolator within the shield, we observed a reduction in isolation performance to \SI{18}{\decibel}, attributed to magnetic feedback within the confined field. In contrast, before the integration within the shield, the measured isolation exceeded \SI{30}{dB}. Despite this degradation, the laser unit remained compatible with our mission requirements.

\begin{figure}
    \centering
    \includegraphics[width=0.79\linewidth]{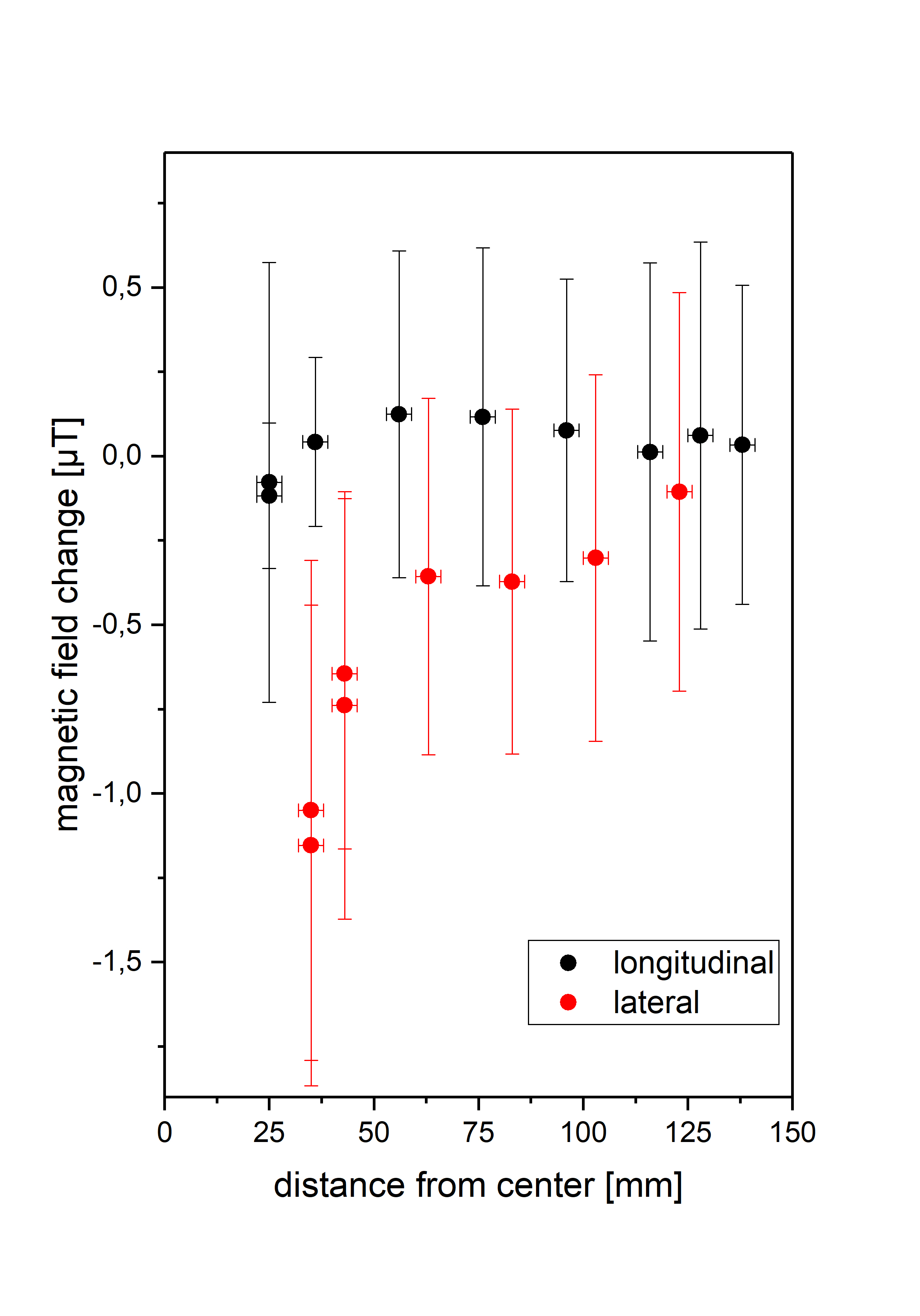}
    \caption{Measured magnetic field at shielded optical isolator as the change compared to the background earth magnetic field}
    \label{fig:Field_Isolator}
\end{figure}

To verify the effectiveness of the shielding, we measured the residual magnetic field using an optically pumped magnetometer (OPM)~\cite{magnetometer} with a sensitivity of \SI{40}{\pico\tesla}. Figure~\ref{fig:Field_Isolator} shows the differential field strength with respect to the ambient geomagnetic background. In the longitudinal direction, no significant deviation was detected within the measurement uncertainty. In the lateral direction, a field change of approximately \SI{1}{\micro\tesla} was observed, consistent with the simulation results. The measurement uncertainty was substantial, primarily because of the unshielded laboratory conditions and ambient magnetic fluctuations from nearby equipment.

The shielded micro-isolator was integrated into the EQM laser unit. Following the isolator replacement, realignment of one coupling mirror was required to recover the optimal fiber coupling. Compared to the configuration with the commercial isolator, the only notable difference is a reduction in output power, attributed to the higher insertion loss of the custom device.

\section{Conclusion and outlook}

In this work, we presented the development of a laser unit operating at \SI{698}{\nano\meter}, designed for space applications and based on off-the-shelf components. The \SI{698}{\nano\meter} laser wavelength is promising for several
space-based applications, including precision spectroscopy,
quantum applications, and strontium-based spaceborne atomic clocks.
Here, the system is specifically tailored to meet the constraints of a 3U Cube-Sat platform for excitation of fluorescent quantum light sources. 
The optical components were individually tested under radiation exposure, thermal cycling, and mechanical stress representative of a low Earth orbit environment. We also detail the assembly process of the optical, mechanical, and electrical subsystems, all engineered to comply with CubeSat integration standards. It is important to note that, since the components were not designed for space applications, our objective is to assess their suitability with the operational requirements of our mission~\cite{Ahmadi_2024}, rather than to demonstrate absolute immunity to degradation.

\begin{figure}
    \centering
    \includegraphics[width=0.8\linewidth]{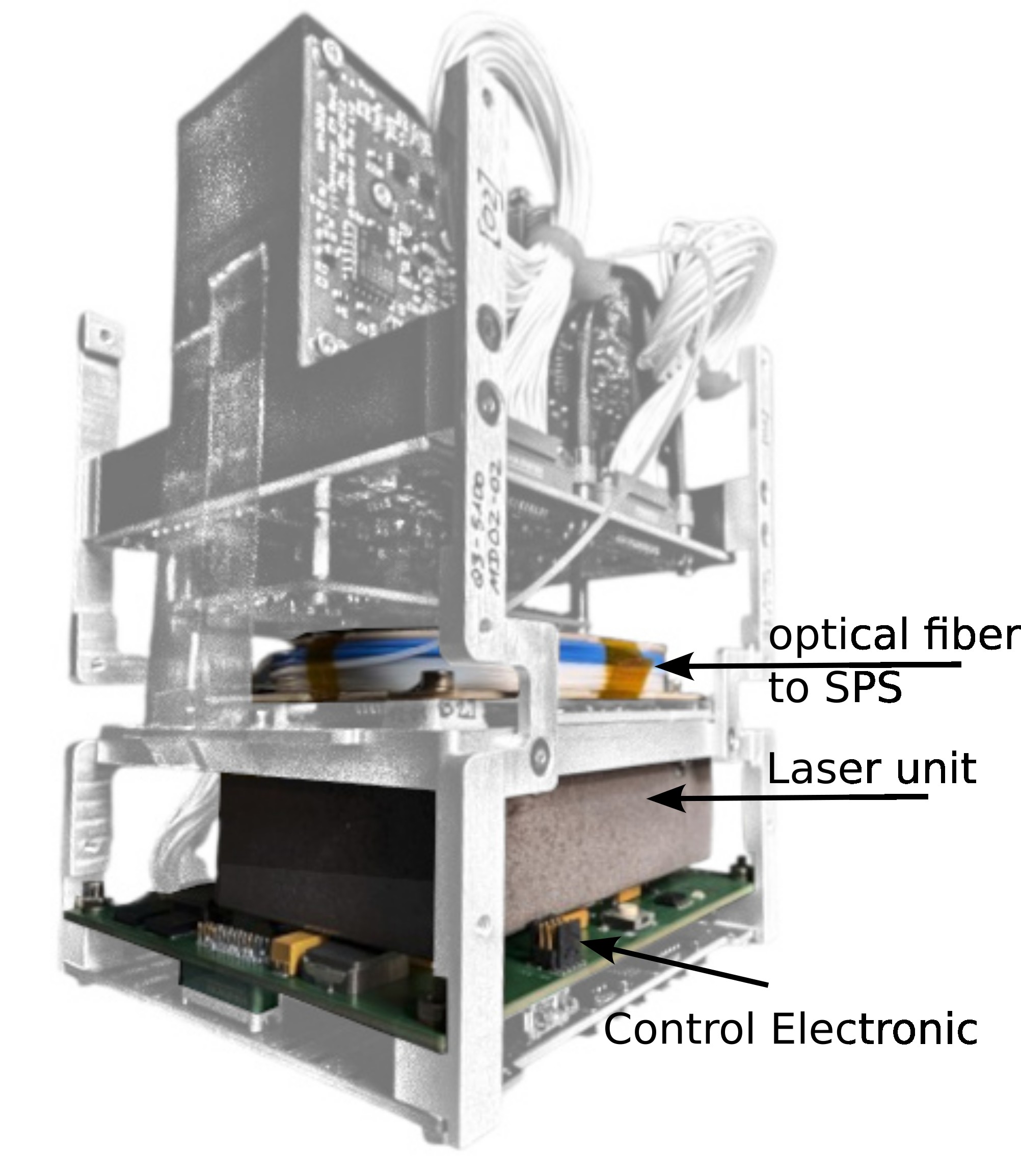}
    \caption{Complete payload: the colored parts indicate the laser unit and driver electronics contributed by the authors. The greyed-out sections show the experiments and the basic structure, which were developed by our QUICK$^{3}$ partners~\cite{Quick3HP} at the Technische Universit\"{a}t Berlin and the Technische Universit\"{a}t M\"{u}nchen (Image courtesy of L. Wiese\textsuperscript{\textcopyright},  TU Berlin).}
    \label{fig:PayLoad}
\end{figure}

As a next step, the fully integrated satellite system will undergo the same qualification procedures applied to individual components, with a focus on verifying subsystem compatibility and interaction. The design and integration of the entire payload shown in Fig.~\ref{fig:PayLoad} were carried out by the QUICK$^{3}$ team at the Technische Universität Berlin. The colored elements were supplied by the authors. Light is transmitted from the laser unit via the blue fiber to the white fiber, which is connected to the SPS. Although the magnetic shielding of the micro-isolator introduces a reduction in optical isolation, it effectively minimizes magnetic interference, an important trade-off also for future magnetically sensitive payloads. Future iterations may benefit from refined shield geometries or alternative materials to reduce these losses.

\section{Acknowledgements}
\noindent The author thanks K. Paschke and G. Blume for providing access to their measurement stations to investigate the laser diodes, and S. Neinert for assistance with the measurements using the optically pumped magnetometer. S.S. thanks M. Christ and C. Zimmermann for fruitful discussions and technical advice regarding micro-integration. S.S. thanks the workshop technicians at FBH for their support and the insight provided during the design phase. 
This work is supported by the German Space Agency
(DLR) with funds provided by the BMWK under grant number No.50WM2166. 
\bibliographystyle{elsarticle-num}
\bibliography{Arxiv/Bib_arxiv}

\end{document}